%% file: paper.tex
\documentclass[conference]{IEEEtran}

\usepackage[pdftex]{graphicx}
\usepackage[utf8]{inputenc}
\usepackage[T1]{fontenc}
\usepackage{microtype}
\usepackage{colortbl}
\usepackage{xargs}
\usepackage{lipsum}
\usepackage{xparse}
\usepackage{xifthen, xstring}
\usepackage{xspace}
\usepackage{marginnote}
\usepackage{etoolbox}
\usepackage{algorithm}
\usepackage[noend]{algpseudocode}
\usepackage{verbatim}
\usepackage{amsmath}
\usepackage{url}
\usepackage{booktabs}
\usepackage[textsize=tiny]{todonotes}

\newcommand{\para}[1]{\par\smallskip\noindent\textit{#1.}}

\newcommand{\specObs}{\phi_{obs_n}}
\newcommand{\specMin}{\phi_{\delta-eq}}
\newcommand{\specBigObs}{\phi_{obs_N}}
\newcommand{\specInt}{\phi_{int-eq}}

\begin{document}

\input{contents/0_titlepage}
\input{contents/1_intro}
\input{contents/2_example}
\input{contents/3_mlir}
\input{contents/4_method}
\input{contents/5_eval}
\input{contents/6_related}
\input{contents/7_conclusion}

\bibliographystyle{plain}
\bibliography{references}

\end{document}

%% file: contents/0_titlepage.tex
\title{mlirSynth: Automatic, Retargetable
Program Raising in Multi-Level IR using 
Program Synthesis}

\author{
  \IEEEauthorblockN{Alexander Brauckmann}
  \IEEEauthorblockA{
    University of Edinburgh \\
    United Kingdom \\
    alexander.brauckmann@ed.ac.uk}
\and
  \IEEEauthorblockN{Elizabeth Polgreen}
  \IEEEauthorblockA{
    University of Edinburgh \\
    United Kingdom \\
    elizabeth.polgreen@ed.ac.uk}
\and
  \IEEEauthorblockN{Tobias Grosser}
  \IEEEauthorblockA{
    University of Edinburgh \\
    United Kingdom \\
    tobias.grosser@ed.ac.uk}
\and
  \IEEEauthorblockN{Michael F. P. O'Boyle}
  \IEEEauthorblockA{
    University of Edinburgh \\
    United Kingdom \\
    mob@inf.ed.ac.uk}
}

\maketitle

\begin{abstract}
MLIR is an emerging compiler infrastructure for modern hardware, but existing programs cannot take advantage of MLIR's high-performance compilation if they are described in lower-level general purpose languages. 
Consequently, to avoid programs needing to be rewritten manually, this has led to efforts to automatically raise lower-level to higher-level dialects in MLIR. However, current methods rely on manually-defined raising rules, which limit their applicability and make them challenging to maintain as MLIR dialects evolve.

We present \emph{mlirSynth} -- a novel approach which translates programs from lower-level MLIR dialects to high-level ones without manually defined rules. Instead, it uses available dialect definitions to construct a program space and searches it effectively using type constraints and equivalences.
We demonstrate its effectiveness by raising C programs to two distinct high-level MLIR dialects, which enables us to use existing high-level dialect specific compilation flows.
On Polybench, we show a greater coverage than previous approaches, resulting in geomean speedups of 2.5x (Intel) and 3.4x (AMD) over state-of-the-art compilation flows for the C programming language. mlirSynth also enables retargetability to domain-specific accelerators, resulting in a geomean speedup of 21.6x on a TPU.
\end{abstract}

%% file: contents/1_intro.tex
\input{figures/motivating_example}

\section{Introduction}
The end of Dennard scaling has led, in recent years, to  the development  of a diverse range of    specialized hardware. Examples include tensor cores (TPU \cite{jouppi2017datacenter} , NVIDIA  \cite{markidis2018nvidia}) and AI specialized accelerators  \cite{reuther2020survey}. Such hardware holds the promise of efficient performance, but at the cost of increased programming complexity.  

A popular approach to overcoming this challenge is the use of domain-specific programming languages  (DSLs), such as  Halide \cite{ragan2013halide}, TensorFlow \cite{abadi2016tensorflow} and  PyTorch \cite{paszke2019pytorch}. 

These languages  allow programmers to easily  specify the essential structure of a problem without concern for low-level details.  Crucially, this separation of concerns enables  domain-specific compilers \cite{wen2021enabling,chen2018tvm}  to efficiently  map programs  down to a wide range of  idiosyncratic   accelerators.

The need for existing code to harness the power of domain specific compilation has been well recognized by the compiler community with the development of MLIR \cite{ lattner2021mlir}.  MLIR is a new extensible representation within LLVM  that captures high-level representations of programs. Once programs are expressed in the appropriate MLIR dialect,  vendors can develop and exploit an  efficient compilation path to  their  platform.

\para{Need for Raising IRs} This, however,  presents a new challenge: how to translate  existing code, currently represented  in a low-level intermediate representation (IR), into a higher dialect so as to leverage domain-specific compilation. 

Furthermore, given the  proliferation of MLIR dialects, any compiler technique that attempts to translate from a low to high-level dialect faces an additional challenge: how to adapt to a world of ever-changing high-level targets.

\para{Previous Approaches}
This lifting of abstraction from a low to a high level is called program  lifting \cite{kamil2016verified} or raising \cite{abadi2016tensorflow,yadavalli2019raising}. There is a large body of work in this area that has  recently received increased interest \cite{mendis2015helium,angstadt2020accelerating,nye2019learning,alur2013syntax,dasgupta2020scalable,collie2019type, hasabnis2016lifting}. In \cite{kamil2016verified}, lifting is applied to legacy FORTRAN code to generate high-performance Halide programs. This is achieved by representing both source and target languages in a common  internal language and deploying an off-the-shelf program synthesis tool \cite{solar2009sketching, torlak2013growing}, and proving the synthesized target code is equivalent to the original  via loop invariants. This was later applied to C++ \cite{ahmad2019automatically} and  expanded into an LLVM-based framework \cite{metalift}. While powerful, such an approach requires the user to manually define the semantics of all operations in the target language semantics in the common internal language. As the number of target high-level languages diversifies, this is not a scalable approach.   
There has been adjacent work in replacing code with  library calls \cite{de2021kernelfarer, martinez2023matching}. Such   approaches, however,  are also fundamentally  non-scalable as they focus on a fixed API rather than the open-ended nature of DSLs and their IRs.

Multi Level Tactics (MLT) \cite{chelini2021progressive}  more recently, directly addressed this issue by showing raising to a high-level MLIR dialect enabled significant performance improvement. However, their approach requires domain-specific raising rules to be implemented by the compiler writer which are  dialect specific. As we show in section \ref{lab:evaluation}, they are  restricted in the number of programs they can tackle and rules need to be rewritten for each source and target dialect. Ideally, we would like a generic scheme that is robust, raising  a large number of programs  and is able to target new MLIR dialects without any compiler writer intervention. 

\para{Our Approach}
This paper presents mlirSynth, which automatically raises  MLIR dialects from low to high-level  without any hardwired compiler transformation or raising rules.   Instead, mlirSynth automatically uses the available dialect definitions (within MLIR's TableGen \cite{lattner2021mlir}) to construct a program  space and effectively searches it using candidate equivalences.  It is based on  bottom-up  enumerative program synthesis exploiting  type constraints and IO behavioral equivalence to quickly prune the space. Essentially it generates programs in the target dialect, starting from the smallest one first. Then it uses a combination of testing and model checking to identify an equivalent program in the target dialect. A key characteristic of our approach is that it is not tied to one dialect. However, if there is domain specific analysis available, we can use this as a heuristic to speed up the search. Thus we have a domain agnostic raiser that can exploit domain specific analysis where available. 

We demonstrate this by lifting to two dialects, Linalg IR and HLO IR, exploiting polyhedral analysis in the synthesis phase.  Furthermore,   we show that our approach is able to cover a wider set of programs and generate more efficient implementations than MLT \cite{chelini2021progressive}, the state-of-the-art scheme. 

Of the 14 Polybench that can be expressed in these dialects, we are able to raise 13, compared to MLT's 6. By exploiting a dialect specific compiler, we are able to achieve an average 20.8x (20.9x) speedup on an Intel (AMD) platform relative to LLVM-O3. This compares to 3.3x (4.2x) of MLT, 6.4x (6.1x) of Polly \cite{grosser2011polly} and 8.1x (4.4x) of Polygeist \cite{moses2021polygeist}. This results in mlirSynth's speedup of 2.5x (3.4x) over these state-of-the-art compilation flows. As mlirSynth can raise to HLO, we can also use the XLA compiler to target Tensor Processing Units (TPUs), achieving a 175.3x average speedup. 

Our contributions are:
\begin{itemize}
  \item mlirSynth, a framework for raising low-level MLIR dialects to higher ones
  \item A scalable method to synthesize code in  multiple MLIR dialects, automatically generating the search space based on the dialect's TableGen definition
  \item A fast bottom-up enumerative search synthesizer exploiting observational equivalence and polyhedral analysis
  \item Greater coverage, performance and accuracy compared to state-of-the-art raising approaches
\end{itemize}

\footnotetext{CPU: Intel I7-8700k (6 cores / 12 threads), TPU: Google TPU v3 (8 cores).}

%% file: figures/motivating_example.tex
\begin{figure*}
    \centering
    \footnotesize
    \begin{minipage}[b]{.6\textwidth}
      \includegraphics[width=1.6\linewidth]{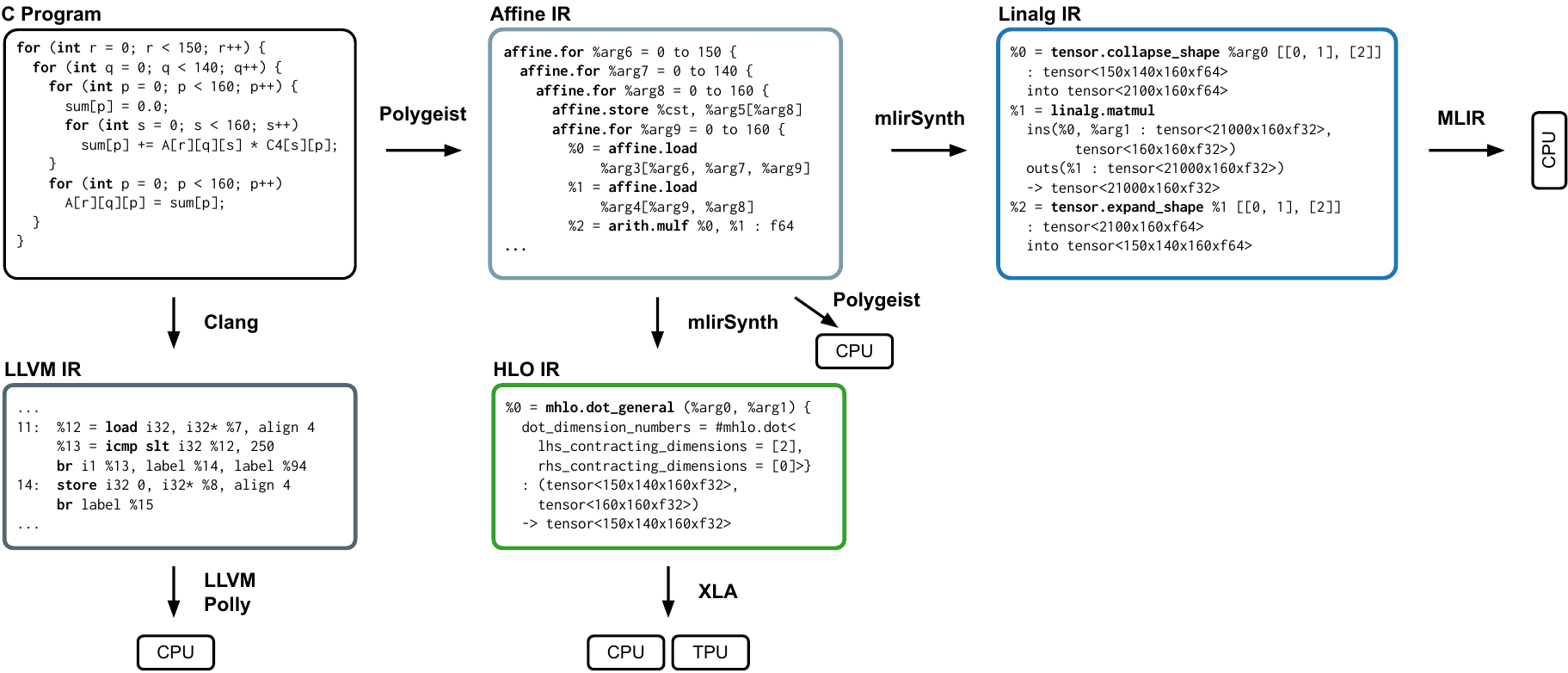}
    \end{minipage}%
    \begin{minipage}[b]{0.4\textwidth}
        \hspace{-0.7cm}
        \vspace{-0.35cm}
      \includegraphics[width=\linewidth]{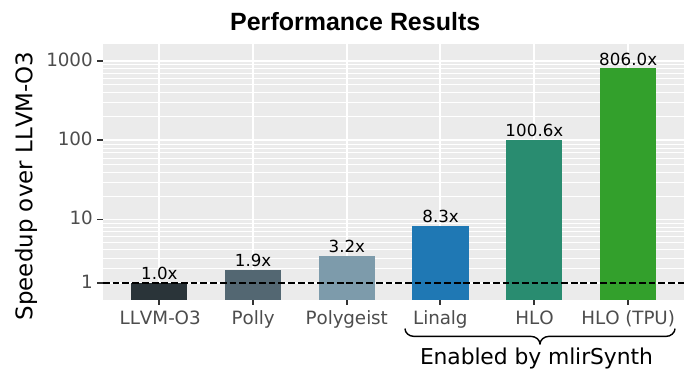}
    \end{minipage}

  \caption{The {\tt doitgen} computation in different representations and their relative performance on different devices.\protect\footnotemark~
  MlirSynth enables compiling the C program with domain-specific compilers such as MLIR-Linalg and XLA, resulting in significant speedups.
  }
  \label{example_xla}
\end{figure*}

%% file: contents/2_example.tex
\input{figures/mlirSynth_dialects_tex}

\section{Motivating Example}

To illustrate the benefits of raising, consider the program in Figure~\ref{example_xla} in the box labeled {\tt C Program}. This loop nest implements the {\tt doitgen} computation and was taken from the Polybench benchmark suite. 

\para{LLVM IR}
As it is written in C, the loop nest is represented within the LLVM compiler by the standard SSA IR form shown in the box labeled {\tt LLVM IR}. If we apply Polly \cite{grosser2011polly}, a polyhedral optimizing compiler to this IR, it is able to automatically generate parallel and cache optimized code. In this case, it is able to achieve a 1.9x speedup over the default -03 pass on an Intel i7 platform as shown in the performance results plot. 

\para{Affine IR}
Although Polly delivers a significant speedup, if the LLVM IR could be rewritten in an alternative MLIR dialect, then we can potentially achieve greater performance. This is the motivation behind the Affine dialect in MLIR, which captures high level polyhedral information, such as linear array access, convex iteration space and static control-flow. Polygeist \cite{moses2021polygeist} is a tool takes in C code and produces Affine IR for appropriate loop nests. They then apply the Pluto \cite{bondhugula2008pluto} polyhedral cache and parallelism optimizer to this Affine IR, which results in a 3.2x speedup. While the performance achieved is greater than Polly, the Affine dialect also acts as a convenient starting point for lifting to higher level dialects and is the source IR for the MLT compiler \cite{chelini2021progressive}. 

\para{Linalg IR}
Consider the Linalg IR version of the program in the box labeled {\tt Linalg IR} in Figure~\ref{example_xla}. It is semantically equivalent to the Affine version, but is in a form that the MLIR compiler can generate more efficient code from. Rather than the Affine IR polyhedral representation of the program, Linalg IR describes the given code as three high level operators: a core matrix multiplication operation, surrounded by two tensor reshaping operations. As matrix multiplication and reshaping are often highly tuned on different hardware targets, it allows the MLIR compiler to exploit kernel libraries and call a highly efficient implementation. The performance results plot shows that if we were able to lift code to this dialect, we were able to achieve an even greater speedup of 8.3x. 

\subsection{Raising IRs}

While MLT tries to automatically raise the Affine code to the Linalg form, it, unfortunately, fails as its hand-coded matching rules do not consider this IR pattern. If additional patterns were added, then it would achieve a higher speedup. 

\para{HLO}
Our approach is not limited to one target dialect of MLIR. It is able to raise to the higher level HLO dialect. Figure~\ref{example_xla} further shows an implementation of the same computation in HLO IR (box labeled {\tt HLO IR}). Rather than the three operations of Linalg IR, the computation is expressed using a single high-level operation. This allows XLA \cite{xla}, a compiler for HLO IR, greater flexibility in its implementation. If we consider the i7 platform, XLA able to achieve a speedup of over 100x relative to LLVM -03, justifying such a dialect. One benefit of this representation is that it can leverage platform specific compilers. If we change the hardware target to a TPU, the XLA compiler delivers a speedup of over 806x. 

\para{Summary}
Higher-level representations in MLIR allow greater performance than lower-level ones and LLVM IR, as they allow compilation with high-performance domain-specific compilers. To enable such compilers for programs written in lower-level programming languages such as C, a raising technique is required. With the ongoing emergence of new dialects, we need a flexible raising technique to automatically leverage high-performance compilation flows. 

%% file: figures/mlirSynth_dialects_tex.tex
\begin{figure*}[t!]
  \centering
  \includegraphics[width=\textwidth]{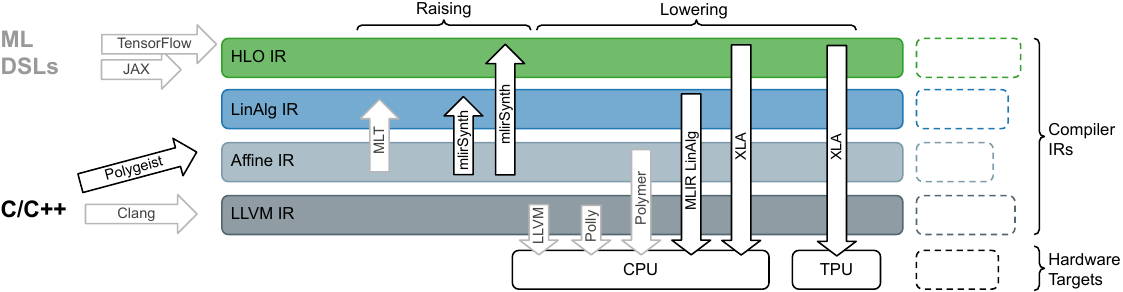}
  \caption{
  Compilation flows relevant in this paper, associating programming languages, MLIR dialects, raising methods, compilers and hardware targets. Flows relevant to mlirSynth are highlighted.}
  \label{fig:dialects}
\end{figure*}

%% file: contents/3_mlir.tex
\input{figures/mlirsynth_internals_tex}

\section{System Overview}
In this section, we briefly introduce the notion of dialects within MLIR before describing the design of mlirSynth's, which raises low-level dialects to higher ones. 

\subsection{MLIR}
MLIR is an infrastructure for developing domain-specific compilers. To aid this, MLIR provides reusable building blocks and shared tools that allow us to define domain-specific languages and their compilation pipelines. The key concept that enables this is a \emph{dialect}. 

Dialects define sets of operations, types, and attributes. There are many dialects currently deployed (35 in the MLIR repository) and, crucially for automatic synthesis, each of these is defined by a structured TableGen description which contains the typed operands, attributes, and regions for each operation. 

Figure \ref{fig:dialects} shows a small subset of MLIR's dialects, relevant to this paper and the associated compilation flows. The lowest level IR we consider is LLVM IR, the default format for languages such as C/C++/FORTRAN. From this, the LLVM compiler generates code for all supported platforms. Many higher dialect compilers can progressively lower their dialect to LLVM IR. 

Different compilation flows exploit high level dialect information to generate efficient implementations. For instance, Polygeist leverages the polyhedral representation available in Affine IR \cite{moses2021polygeist}. While some programming languages / DSLs such as TensorFlow can benefit from the XLA compiler due to its representation as HLO, this is not available to languages such as C. If we can raise code using MLT or mlirSynth to higher MLIR levels, then we can leverage the pre-existing compilation flows for performance. While a lowering path is always provided, raising is considerably more challenging. 

\subsection{mlirSynth design}

mlirSynth operates in a three stage process that takes in a user program in the Affine IR dialect plus a description of the target dialect and outputs a raised program in the new dialect, as shown in Figure \ref{fig:mlirsynth_internals}. The central idea is that we apply classic program synthesis techniques to lift a dialect to a higher one and then lower both the original and raised program down to the same representation and check they are equivalent. We use smallest-program first enumeration, discarding candidate programs based on type information and guided by heuristics where available. The approach is comprised of 3 stages: 

 \para{Pre-processing} Initially, we apply pre-processing steps to simplify the program before enumerative synthesis is applied. Specifically, we use polyhedral analysis to distribute the original loop structure into smaller ones that can be synthesized independently and reduce the size of any array/matrix-like data structures in the code. 

 \para{Enumerative synthesis} The key lifting is done by a classic enumerative synthesis technique. We generate a grammar (box {\tt Grammar Generator}) for the synthesis process from the target dialect's TableGen definition ({\tt Target Dialects}), automatically generate an input-output specification ({\tt IO Spec Generator}) for the target program, and then use a bottom-up synthesis process ({\tt Enumerator}) to search the space of possible programs. This space is large, so we guide the enumerative search with a number of heuristics, based on polyhedral analysis and observational equivalence. Lifted candidates that satisfy the input-output specification ({\tt Checker}) can be proved to be equivalent to the original code using bounded model checking ({\tt Validator}). 

 \para{Post-processing} Once we have a successful candidate, it is inlined back into the program and all data structure sizes are restored. It is now in the appropriate high level dialect and can be mapped to hardware by an appropriate compiler toolchain.

\smallskip

The following section gives details of the enumerative synthesis stage.

%% file: figures/mlirsynth_internals_tex.tex
\begin{figure*}[t!]
  \centering
  \includegraphics[width=\textwidth]{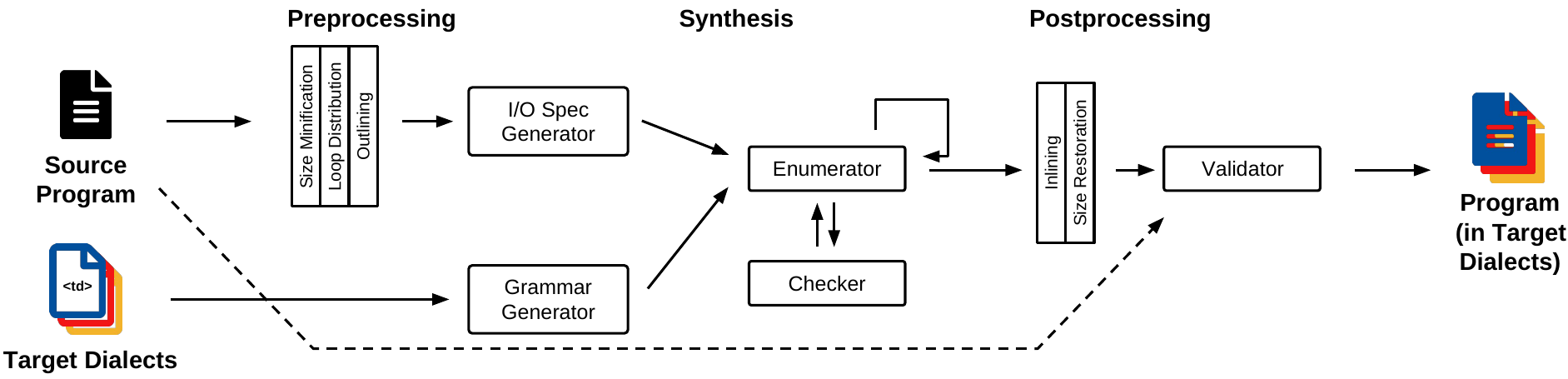}
  \caption{Design of mlirSynth, showing its processing pipeline and component interaction.}
  \label{fig:mlirsynth_internals}
\end{figure*}

%% file: contents/4_method.tex
\section{Synthesis}
\label{sec:synthesis}

We base our synthesis procedure on classic enumerative synthesis algorithms from the literature~\cite{barke2020just}. The core synthesis loop, shown in Algorithm~\ref{alg:synth}, is two phases: enumerating candidate programs and then checking the candidate against a specification. We thus need two inputs: a grammar that defines the space of possible programs to enumerate and a specification for the synthesized program. In the following section, we describe how we automatically generate grammars and specifications for our synthesis algorithm, the exact details of the enumeration process and the heuristics we use. 

\subsection{Grammar generation}
In MLIR, dialects are defined in the declarative TableGen language. This language provides a structured way to define new dialects, including specifying the number and types of operands, attributes and regions that are required per operation. Given a target MLIR dialect, we use a custom TableGen extraction tool to generate a simple recursive context-free grammar for our enumerative synthesis process. Specifically, we generate an initial grammar $G$ that contains a single non-terminal for each type in the MLIR dialect, and a production rule for each operator. 

This grammar does not precisely capture the syntactic restrictions of the MLIR dialect (since MLIR dialects are in general not context-free languages), but we are able to use inbuilt MLIR syntactic checks to discard any invalid programs later on in the enumerative loop. The grammar is, however, specific enough to rule out the majority of poorly typed programs. This automatic grammar generation simplifies retargeting our technique to new dialects. 

\subsection{Specification generation}
Given a grammar $G$ and a reference function $f$ that we wish to lift, which takes input $x$, our goal is to synthesize a function $f'$, such that $\forall x. f(x) = f'(x)$ and $f'$ and $f'$ is in $\mathcal{L}(G)$, the language defined by the grammar. 

Checking a function satisfies the full equivalence specification above, for arbitrarily large input data structures, is in general undecidable and a significant challenge to state-of-the-art verification tools. Consequently, we apply a multi-staged synthesis process using two approximations of the specification. In the first specification, $\specMin$, we minify the input data structures, reducing $x$, a potentially very large data structure, to $x_{min}$, a small, bounded size data structure. We then assert that the relative error between the outputs of the functions is smaller than a small $\delta$. This $\delta$ accounts for the deviations introduced by compiler optimizations due to non-associativity of floating-point arithmetic. The second specification checks the observational equivalence of the functions, i.e., it checks for equivalent behavior on a finite set of $n$ inputs. We denote this $\specObs$ when it checks behavior on $n$ inputs and $n$ is small (i.e., less than $10$) and $\specBigObs$ when it checks behavior on $N$ inputs, where $N$ is large (i.e., $\geq 10$). Formally, the specifications hold for a given candidate $f'$ under the following conditions: 

\begin{align*}
\specMin\,\, & \Leftrightarrow \forall x_{min}. abs(f(x_{min}) - f'(x_{min}))/f(x_{min}) < \delta\\
\specObs\, & \Leftrightarrow \forall i\in I_n. f(i) = f'(i) \\
 &\text{where $I_n$ is a small finite set of $n$ inputs}\\
\specBigObs & \Leftrightarrow \forall i\in I_N. f(i) = f'(i) \\
 &\text{where $I_N$ is a large finite set of $N$ inputs}\\
\end{align*}

We automatically generate $\specObs$ and $\specBigObs$ by randomly generating sets of inputs. We then check for observational equivalence by compiling and executing both $f$ and $f'$ on the inputs, shown in Algorithm~\ref{alg:speccheck}. The data for $\specObs$ is initially sampled from the range $[-10, 10]$ to work around the effect of numerical instabilities and to avoid cases where it is likely that the synthesized code will crash. 

For speed, the core enumerative algorithm, Algorithm~\ref{alg:enum} checks candidates against $\specObs$. Initially, $I_n$ contains only a single input example, although this input is a high-dimensional tensor or matrix. The outer loop checks candidates against $\specBigObs$. We use CBMC~\cite{cbmc}, a bounded model checker, to check $\specMin$ post-synthesis. 

\input{figures/mlirSynth_synthesis_example_tex}
\input{algorithms/enumerate}

\subsection{Bottom-Up Enumeration}
The core enumeration is a bottom-up synthesis algorithm inspired by~\cite{AlbarghouthiGK13}. The enumeration combines previous candidates with each other to generate more complex ones until a candidate matching the specification is found. An example of this is shown in Figure~\ref{fig:example}. 

\para{Initialisation}
We start by creating a candidate set $C$ of valid (i.e., well-formed) candidates that each produce a computationally distinct value from the other candidates. We initialize this set with candidates that return the arguments of the reference function $f$, as shown in the left-most \texttt{Candidate Set} box in Figure~\ref{fig:example}, as well as simple constants in the shape and data type of the arguments and results of the function. We will use this active candidate set as the base of our enumeration. 

\para{Enumeration}
The synthesis loop enumerates through the set of operations in the grammar. For each operation, we first identify sets of possible operands, attributes and regions. We do this according to the operation signature in the grammar. 

We populate the set of possible operands with all expressions in the candidate set of the correct type, highlighted yellow in each iteration in Fig~\ref{fig:example}. For operators with $y$ operands of type $\tau$, we add each active candidate of type $\tau$ to the set of possible operands $y$-times to allow operators to have two or more identical operands. 

For the attributes, which need to be known statically, we generate a large number of them, depending on their type, once the operator is selected. 
For regions, which contain groups of operations with arguments, we generate simple ones that perform binary mathematical operations on the function arguments (specifically, the operations addition, subtraction, multiplication and division). Region generation is not shown in Figure~\ref{fig:example}, but these are generated in each iteration after the operator is selected.

We generate a set of all possible candidates by taking the Cartesian product of sets of operands, attributes and regions. 

\para{Candidate Checking}
Each candidate in the set is validated using a series of static checks, ordered by their complexity, such that the cheapest checks are performed first, and the expensive checks are performed last. We use MLIR's type and shape inference system and built-in verification method chain to perform these checks. 

\para{Equivalence pruning and validating the candidate}
If the static checks succeeded so far, we use MLIR's execution engine to just-in-time compile the candidate. We then check $\specObs$ by executing the candidate program $f'$ on the set of inputs and comparing the output value with the output value produced by the reference function. If $\specObs$ is satisfied, we send the candidate to the outer loop check. 

We also check if the candidate is observationally unique on the inputs used by $specObs$, that is, there does not exist a candidate in the candidate set $C$ that behaves the same as $f'$ on all inputs. In other words, if $f'$ is observationally unique, the following formula is valid $\not\exists f_c \in C. \forall i \in I_n f'(i)=f_c(i)$. If this is the case, we add $f'$ to the candidate set $C$. If not, we discard the candidate. 

This central enumeration process is repeated for each operation, until either a program matching the specification $\specObs$ is found, or a timeout expires. Once a candidate satisfies $\specObs$, it is then checked against $\specBigObs$. If a candidate fails this check, we can restart the synthesis loop with a new set of random inputs for $\specObs$. 

\subsection{Heuristics}
Given the search space, heuristics are essential for selecting a set of operations to enumerate. We implement two heuristics: polyhedral model-based and dialect-based heuristics. These heuristics alter the behavior of the function \texttt{pickOperations}. Polyhedral model-based heuristics perform a value-based dependence analysis on the reference function to identify reduction dependencies, which are visible as cycles in the polyhedral dependence graph. If such reduction dependencies exist, this heuristic selects reduction operations in the target dialect. For a more detailed discussion of this see \cite{doerfert2015polly,yang2021simplifying} 

Dialect-based heuristics look at the grammar for the target dialect and the source dialect, and, if an operator is present in the source function, written in the source dialect, it is prioritized in the target dialect. For example, if an add operation exists in the function, this heuristic selects any arithmetic operations that perform an addition in the target dialect. We evaluate the impact of heuristics in Section \ref{sec:synthesis_time}. 

\subsection{Translation Validation}
\label{sec:finalcheck}
We use CBMC~\cite{cbmc} to perform a post-synthesis check for the specification $\specMin$. CBMC uses symbolic execution to generate a logical formula that is satisfiable if and only if the two functions are not equivalent (or, in path-based mode, multiple logical formulas representing different paths through the program). If the functions are not equivalent, CBMC will generate a \emph{counterexample}, in the form of a set of inputs for which the outputs of the two functions are not identical. If this were to happen, we could repeat the synthesis process, using the counterexample input generated as one of the input examples in specification $\specObs$. If CBMC exceeds a timeout of $1$ hour, we substitute this check with extensive testing of $\specBigObs$. 

%% file: figures/mlirSynth_synthesis_example_tex.tex
\begin{figure*}
  \centering
  \includegraphics[width=\linewidth]{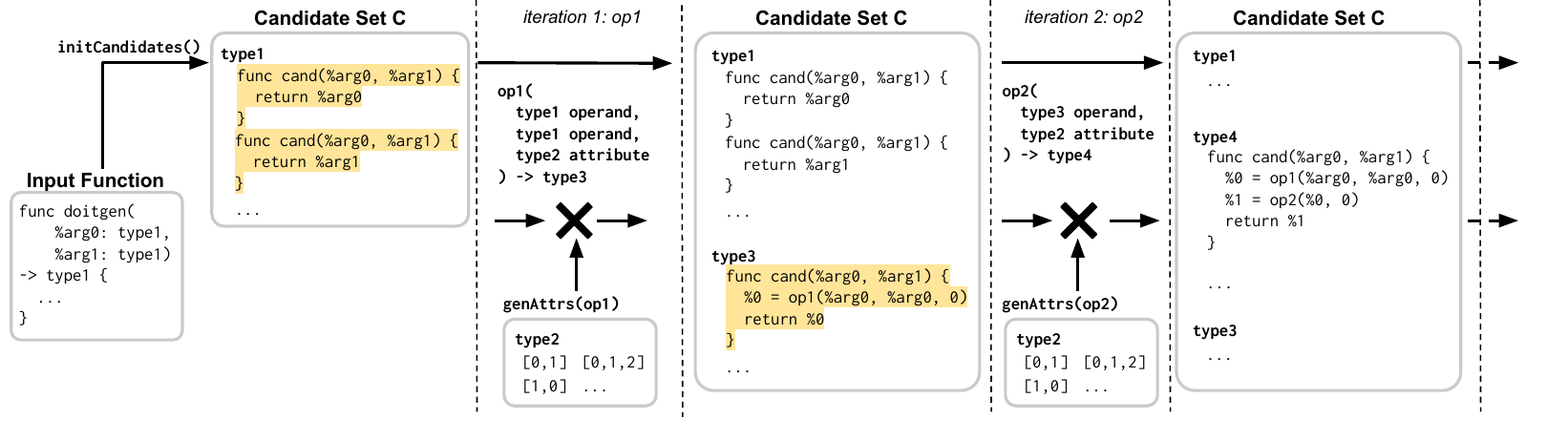}
  \caption{Synthesis example showing the enumeration of the candidate set at each iteration. In each iteration, an operator is selected, and the active set (highlighted in yellow) is chosen based on the types of the operands, and a set of attributes and regions is generated. We enumerate through the cartesian product until a correct candidate is found. If no correct candidate is found, any observationally unique candidates are added to the candidate set and we start a new iteration with a new operator. }
  \label{fig:synthesis_example}
  \label{fig:example}
\end{figure*}

%% file: algorithms/enumerate.tex
\begin{algorithm}[t!]
  \begin{algorithmic}
    \Function{synthesize}{$f$,$G$}
    \State $C \gets \text{initCandidates}(f)$
    \State $I_n \gets \text{genRandomInputs}(f, n)$
    \State $operations \gets \text{pickOperations}(f,G)$
    \While{true}
      \State $f' \gets \text{enumerate}(C, I_n, operations, f)$
      \State $I_N \gets \text{genRandomInputs}(f, N)$
      \If{$\text{specCheck}(I_N,f,f')$}
      \State \textbf{return} $f'$
      \Else
      \State $I_n \gets \text{genRandomInputs}(f, n)$ %
      \EndIf
      \EndWhile
   \EndFunction
  \end{algorithmic}
  \caption{Core synthesis algorithm\label{alg:synth}}
\end{algorithm}

\begin{algorithm}
  \begin{algorithmic}
    \Function{enumerate}{$C$, $I_n$, $operations$, $f$}
    \While{true}
    \For{$op$ \textbf{in} $operations$}
    \State $ops \gets \text{filterTypes}(C, op)$
    \State $attr \gets \text{genAttrs}(op)$
    \State $regs \gets \text{genRegions}(op)$
    \For{$f'$\textbf{in} \text{cartesianProduct}($ops$, $attr$, $regs$)}
    \If{not staticCheck($f'$)}
    \State \textbf{continue}
    \EndIf
    \If{$\text{observationallyUnique}(C,f')$}
    \State $C \gets C \cup f'$
    \EndIf
     \If{$\text{specCheck}(I_n,f, f')$} %
     \State \textbf{return} $f'$
     \EndIf
    \EndFor
     \EndFor
    \EndWhile
    \EndFunction
  
  \end{algorithmic}
  \caption{Enumeration\label{alg:enum}}
\end{algorithm}

\begin{algorithm}
  \begin{algorithmic}
  
  \Function{specCheck}{$I$, $f$,$f'$ }
  \For{$i$ \textbf{in} $I$}
  \If{$f(i) \neq f'(i)$}
    \State \textbf{return} $false$ %
  \EndIf
  \EndFor
  \State \textbf{return} $true$ %
  \EndFunction
    
  \end{algorithmic}

  \caption{Specification checking\label{alg:speccheck}}
\end{algorithm}

%% file: contents/5_eval.tex
\input{tables/coverage}
\input{plots/overview}
\input{plots/coverage_tex}

\section{Experimental Setup}
This section briefly describes the platforms, benchmarks and various compilation flows used to compare against mlirSynth.

\subsection{Platforms}
All experiments were performed on two multi-core hardware platforms an Intel i7-8700k and an AMD Ryzen 9 3900X with multi-threading enabled. We also evaluated on 1 domain specific accelerator, the Google TPU v3.8. 

\subsection{Benchmarks}
We evaluated all techniques on those benchmarks from Polybench that can be represented in either Linalg and/or HLO. These are shown in Table \ref{table:coverage}. We used Polybench 4.2.1-beta in the large data size and float configuration. 

\subsection{Compilation flows}
We evaluated a number of different compilation flows:\\
{\bf LLVM-O3}: General-purpose compiler, used as baseline \cite{llvm} \\ 
{\bf Polly}: Polyhedral compiler, optimizes LLVM IR for caches and parallelism \cite{grosser2012polly}\\
{\bf Polygeist}: Polyhedral compiler, takes C to Affine IR, then
uses Pluto for cache- and parallelism optimization \cite{moses2021polygeist} \\
{\bf MLT}: Raises Affine to Linalg IR before invoking a tuned MLIR Linalg compilation flow \cite{chelini2021progressive} \\ 
{\bf MLT-BLAS}: As MLT, replacing named Linalg operations with BLAS calls after raising \cite{chelini2021progressive}\\
{\bf mlirSynth-Linalg}: Our approach, raises Affine to Linalg IR (on tensors) and invokes latest untuned MLIR Linalg compiler \\
{\bf mlirSynth-XLA}: Our approach, raises Affine to HLO IR and then invokes the XLA compiler targeting CPUs\\
{\bf mlirSynth-XLA (TPU)}: As mlirSynth-XLA, targeting TPU\\

\subsection{Methodology}
All experiments were run 10 times with median end-to-end execution time reported. To ensure that there was no caching, we ensured a cold start for each experiment, spawning a new process for each run. To further evaluate our raising ability we compare against MLT \cite{chelini2021progressive} and KernelFaRer~\cite{de2021kernelfarer}, which are state-of-the-art methods for raising to high-level operations from lower-level code. 

 mlirSynth-Linalg uses the latest default MLIR lowering compiler without any tuning or replacement of named linalg operations with BLAS calls. MLT, however, uses a legacy version of MLIR tuned for performance (e.g. optimized tile sizes). To provide a fair comparison, we retain MLT's use of this tuned legacy compiler. mlirSynth-Linalg currently targets the exact same subset of Linalg as MLT to allow side-by-side comparison. This, however, restricts the number of programs that can be raised. 

While mlirSynth could raise programs to combinations of target dialects, we are limiting the experiments to individual ones to allow a more tractable search. We plan to explore combinations of target dialects in future work. 

\input{plots/detailed.tex}
\input{plots/synth_time}

\section{Evaluation}
\label{lab:evaluation}
This section first summarises the performance achieved by each approach before examining coverage. This is followed by a detailed performance comparison and an analysis of mlirSynths compilation time. It concludes with an validity evaluation. 

\subsection{Overall Summary}
Figure~\ref{fig:speedups} shows the average speedups of the various compilation flows on three platforms across the Polybench benchmarks. Speedups are relative to LLVM-O3. MLT lifting to Linalg achieves a 1.1x speedup on the AMD platform, and 1.2x on Intel. Replacing named operations with BLAS routines however achieves 3.3x improvement on the Intel platform, rising to 4.2x on AMD. Although significant, MLT-BLAS's performance is limited by the number of kernels it can raise. In fact, Polly is able to achieve greater improvement: 6.4x and 6.1x speedup on each platform as it can optimize more kernels. Polygeist is able to exploit the Affine IR representation, achieving 8.1x and 4.4x speedups. mlirSynth-Linalg is able to synthesize a larger number of kernels than MLT, which gives a performance improvement across both CPU platforms. However, MLT-BLAS uses a tuned MLIR legacy compiler and substituted BLAS routines gives increased performance. When lifting to HLO and invoking the XLA compiler, mlirSynth-XLA achieves a geometric mean speedup of 20.8x on the Intel platform, rising to a geometric mean of 20.9x on the AMD. This significant increase is because XLA makes use of vendor-optimized kernel libraries. When targeting the TPU, mlirSynth is able to achieve over 175x speedup, a significant improvement. 

\subsection{Coverage}
The performance achieved by raising critically depends on the number of raised programs. Figure~\ref{fig:coverage} shows the percentage of programs raised by different approaches for each of the Polybench categories shown in Table \ref{table:coverage}. KernelFarer \cite{de2021kernelfarer} is a robust GEMM detector and is able to detect two routines in the BLAS and kernel category, but no others due to its hardwired pattern matching rules. MLT performs better capturing most of the kernels (5 out of 6) and some of the BLAS (3 out of 7) categories. It is unable to capture any of the data mining kernels due to its restricted Linalg coverage. When raising to HLO, mlirSynth is able to raise all candidates except for {\tt trmm}. Its computation cannot be represented in HLO operations and therefore, results in raising to fail. 

\subsection{Detailed CPU performance results}
Figure~\ref{fig:speedups_detailed} shows a more detailed performance evaluation of different compilation paths relative to LLVM-O3. We show Polly and Polygeist as the polyhedral compilers, MLT, MLT-BLAS, finally mlirSynth raising Linalg and XLA operations. 

MLT-BLAS is able to improve on the polyhedral compilers on the Intel platform in 3 cases where large matrix multiplications dominate execution time, {\em 2mm, 3mmm, gemm}. On the AMD platform, in addition, it performs well on {\tt atax}. Overall however MLT-BLAS performs less well than polyhedral compilers as it is unable to raise all the benchmarks to Linalg. 

As mlirSynth-Linalg is able to synthesize a larger number of kernels than MLT, it achieves performance improvement in seven of the benchmarks. In 2 cases, {\tt bicg} and {\tt gemver} MLT performs better due to its tuned legacy MLIR compiler. 

Comparing mlirSynth-XLA to MLT-BLAS, we see MLT-BLAS is able to achieve comparable performance to mlirSynth-XLA on matrix-matrix multiply like kernels ({\tt 2mm, 3mm, gemm}). This is because XLA uses similar BLAS kernel libraries like MLT, particularly on the Intel platform. On other programs such as {\tt mvt}, mlirSynth-XLA is superior as it synthesizes a more efficient, but computationally equivalent program. 

It is clear that raising beyond Linalg to HLO enables significant performance improvement due to the superior XLA compilation flow. On both of the CPU platforms, it outperforms the corresponding Linalg implementations in all cases. While it is less performant than polyhedral compilers for small problems, on average it significantly outperforms them. 

\subsection{TPU results}
If we consider the domain-specific TPU accelerator, then HLO enables even greater performance improvement across the benchmarks, as shown in Figure \ref{fig:TPU}. The XLA compiler is able to achieve over 3000x speedups in some cases, performing particularly well on programs containing large matrix operations. 

\input{tables/synth_stats}

\subsection{Synthesis time}
\label{sec:synthesis_time}

While mlirSynth is able to raise programs to higher-level dialects, it requires search-based synthesis, which could be expensive. Figure~\ref{fig:synth_time} shows the synthesis time for each benchmark for each dialect and compares it against a naive synthesizer that does not restrict the types, set of operations and equivalences. 

In 7 out of 13 cases, we are able to raise programs in less than 2 seconds. In the 4 more complicated examples, synthesis time increases to over 2 minutes. The impact of type information and candidate pruning is significant. Compared to the naive algorithm, we are able to reduce synthesis time by an order of magnitude. 

Table \ref{table:synth_stats} provides a more detailed breakdown of the synthesis process for HLO. The synthesis time is correlated to the largest synthesis subproblem size, whereas the largest solved one is 5 high-level operations. While the number of candidates considered varies from 720 to c. 20 million, over 90\% of these can be discarded based on type and shape filtering. The remaining candidates are then evaluated, with those that are found to be equivalent to existing candidates eliminated from further consideration. The number discarded this way varies and increases in impact as the number of enumerated candidates increases: from less than 1\% for 3mm up to 60\% for symm. 

\subsection{Validity}
\label{sec:validity}
As described in section~\ref{sec:finalcheck}, we employ model checking~\cite{cbmc} to determine if our raised program is equivalent to the original~\cite{cbmc-equiv}. In our post-synthesis checks, 4 solutions are proven to be equivalent or $\delta$-equivalent (satisfying $\specMin$). CBMC failed to find any errors within $1$ hour when checking $\specMin$ for 8 of the remaining HLO solutions, but we were able to prove that these are equivalent when using integers and discounting floating-point arithmetic. That is, the specification $\specInt$ was shown to hold, where $\specInt \Leftrightarrow \forall x\in X_{int}. f_{int}(x)=f'_{int}(x)$ and $X_{int}$ is the set of all possible integers and $f_{int}$ and $f'_{int}$ use integer arithmetic throughout. We are unable to verify the remaining two solutions (correlation and mvt) due to bugs in CBMC but all solutions passed extensive testing, i.e., $\specBigObs$ holds. We identify 2 issues: 

1) All queries use small input data structures so there is a risk that bigger floating-point errors will accumulate on larger data structures. There are also some solutions where CBMC times out and we default to using integer arithmetic, so there is a small risk that an error trace for floating point might exist. These risks are mitigated by testing performed on large numbers of input examples with variable data structure sizes. 

2) The specification $\specMin$ permits a relative error of $\delta=10^{-5}$. $\delta$ is chosen to account for subtle differences introduced by the MLIR and XLA compilers due to the order in which floating-point operations are performed but this results in a precise verification tool like CBMC reporting that the synthesized functions are not equivalent. For practical purposes, we can thus consider the solutions that are proven to satisfy $\specMin$ to be equivalent. 

\subsection{Discussion}
Enumerative search is able to raise programs to two MLIR dialects and leverage the pre-existing compiler flows to deliver portable, high-performance code. In particular, it enables access to accelerators such as TPU from low-level languages. While IO testing in practice is sufficient for correct lifting, verification is needed to guarantee correctness. Existing raising schemes such as MLT provide no such guarantee. If the compiler writer inadvertently inserts an incorrect rule, it is not checked. In fact, we discovered that MLT incorrectly identified an in-place matrix update as a functional operation and actually gave incorrect results in one case. Synthesis times for bottom-up enumeration scales with program complexity. For more complex dialects, smarter sketch-based or probabilistic schemes will be useful and the subject of future work. 

%% file: tables/coverage.tex
\begin{table}
\centering

\caption{Dialect coverage across PolyBench}

\begin{tabular}{ll|ll} 
\toprule
 Category & Benchmark &  Category & Benchmark\\ [0.5ex]
\midrule
    datamining &    correlation & blas & syrk \\
               &     covariance &   &  syr2k  \\
       kernels &           atax &   & trmm    \\
               &            2mm &   & gemver  \\
               &            3mm &   & symm    \\
               &        doitgen &   & gesummv \\
               &            mvt &   & gemm    \\
               &           bicg  &  &         \\
\bottomrule
\end{tabular}
\label{table:coverage}
\end{table}

%% file: plots/overview.tex
\begin{figure*}
    \begin{minipage}[b]{.8565\textwidth}
      \includegraphics[width=\linewidth]{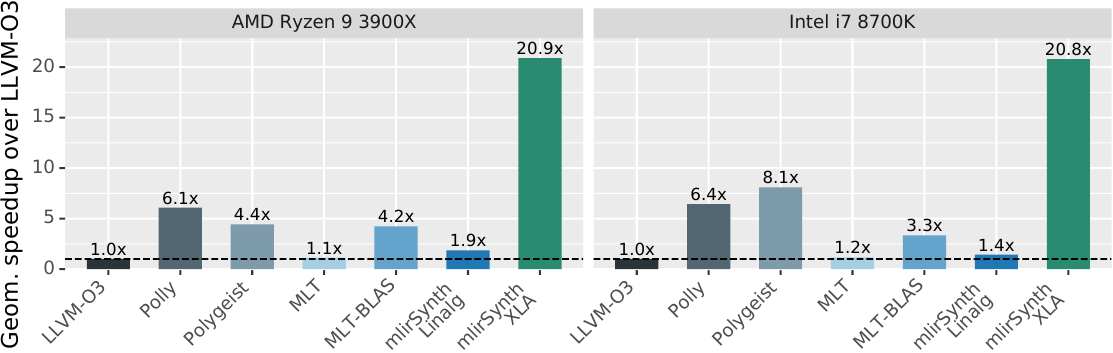}
    \end{minipage}%
    \begin{minipage}[b]{.124\textwidth}
      \includegraphics[width=\linewidth]{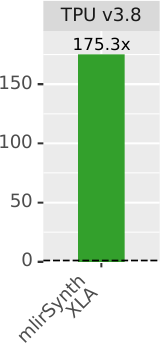}
    \end{minipage}
    
    \caption{Geo-mean speedup of each compilation approach on 3 different hardware platforms: Intel CPU, AMD CPU, and TPU.}
    \label{fig:speedups}
\end{figure*}

%% file: plots/coverage_tex.tex
\begin{figure}
    \includegraphics[width=0.95\linewidth]{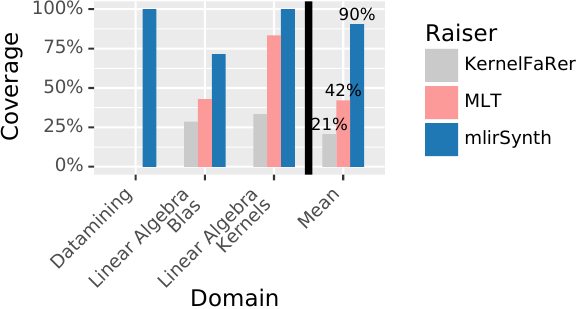}
    \centering
    \caption{Comparison of coverage across Polybench benchmarks using different raising techniques.}
    \vspace{-0.2cm}
    \label{fig:coverage}
\end{figure}

%% file: plots/detailed.tex
\begin{figure*}
    \includegraphics[width=\textwidth]{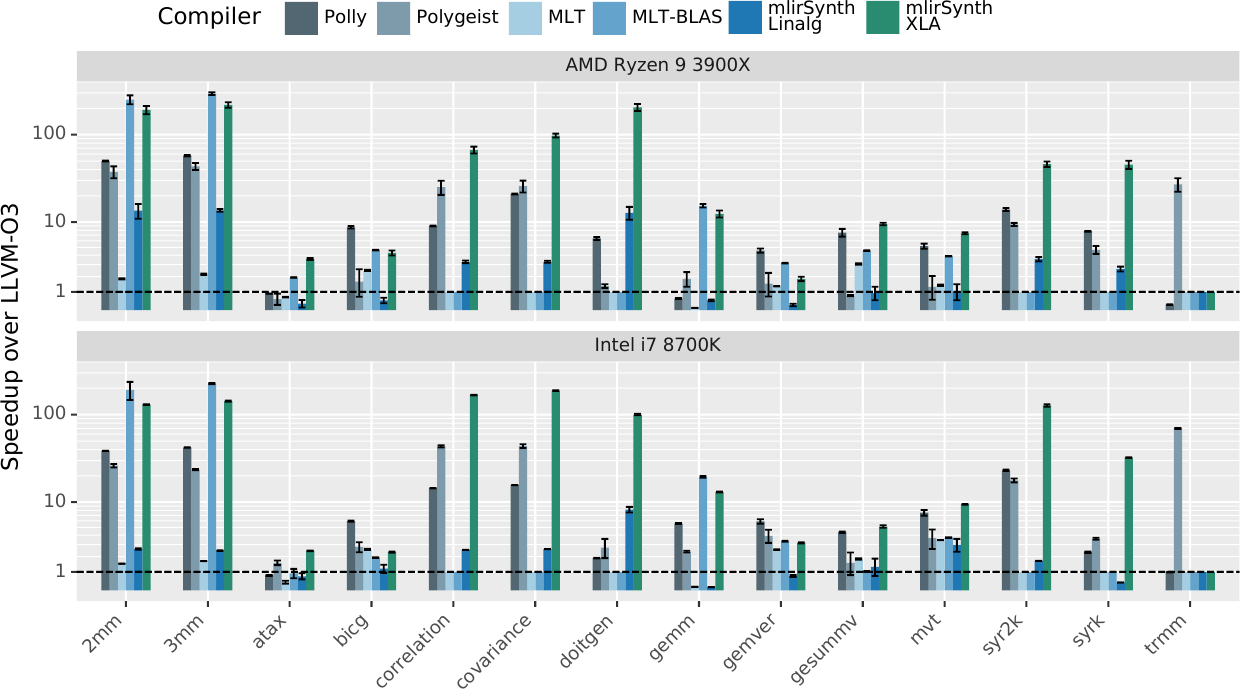}
    \caption{Detailed speedups on the CPU platforms. Bars show median, lines the standard deviation of 10 runs.}
    \label{fig:speedups_detailed}
\end{figure*}

\begin{figure}
    \centering
    \includegraphics[width=0.9\linewidth]{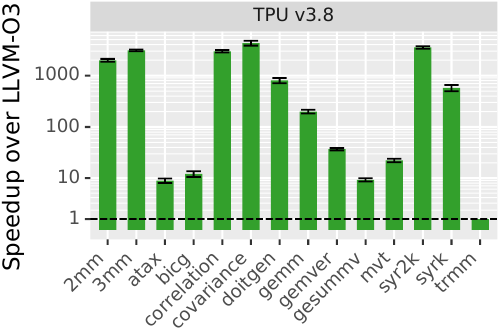}
    \caption{Detailed speedups on TPU over the Intel i7 8700K. Bars show median, lines the standard deviation of 10 runs.}
    \label{fig:TPU}
    \vspace{-0.6cm}
\end{figure}

%% file: plots/synth_time.tex
\begin{figure*}
    \begin{minipage}[b]{.307\textwidth}
      \includegraphics[width=\linewidth]{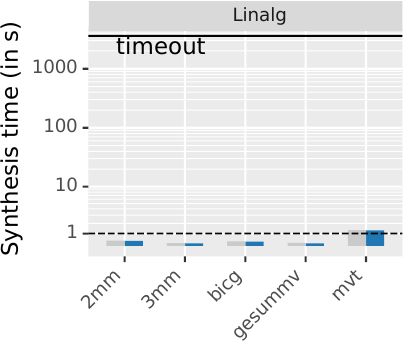}
    \end{minipage}%
    \begin{minipage}[b]{.693\textwidth}
      \includegraphics[width=\linewidth]{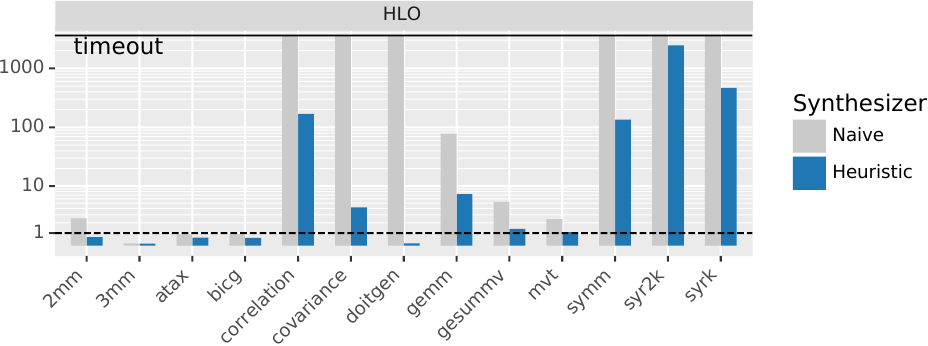}
    \end{minipage}

    \caption{Synthesis times in seconds of different benchmarks in the Linalg and HLO dialects.}
    \label{fig:synth_time}
\end{figure*}

%% file: tables/synth_stats.tex
\begin{table*}
\centering

\caption{Synthesis statistics of mlirSynth on HLO in heuristic mode. Static checks include filtering based  type correctness and additional checks via dialects verification interface. The final column shows the level of post-synthesis guarantees (Section \ref{sec:validity}).}

\begin{tabular}{l|rrrr|rr|r}
\toprule
Benchmark &  Enumerated &  Static filtered &  Evaluated &  Equiv filtered & Ops (max) & Synth time & Formal Guarantee \\
\midrule
        2mm &       49067 &          46504 &       1043 &                   709 &    7 (3) &                0.65s & $\specInt$ \\
        3mm &        2484 &           2409 &          3 &                     0 &    3 (1) &                0.14s & $\specInt$\\
       atax &       18960 &          17042 &       1166 &                   763 &    8 (3) &                0.62s & $\specMin$\\
       bicg &       18961 &          17046 &       1173 &                   771 &    8 (3) &                0.59s & $\specMin$ \\
correlation &     1420241 &        1173035 &     188577 &                159679 &    22 (3)&              174.11s &  $\specBigObs$\\
 covariance &      382674 &         374083 &       5799 &                  2049 &    6 (3) &                4.21s & $\specMin$\\
    doitgen &        9972 &           9879 &         71 &                    18 &    1 (1) &                0.16s & $\specInt$\\
       gemm &      607638 &         572798 &      13695 &                  6745 &    4 (3) &                7.26s & $\specInt$\\
    gesummv &       29221 &          24566 &       3919 &                  2333 &    9 (3) &                1.37s & $\specMin$\\
        mvt &       27977 &          24460 &       2855 &                  1631 &    4 (2) &                1.09s & $\specBigObs$\\
       symm &     5353361 &        4943595 &     309752 &                163310 &    4 (4) &              134.85s & $\specInt$\\
      syr2k &    20820281 &       18547932 &    1467901 &               1022725 &    12 (5)&             2438.69s &  $\specInt$\\
       syrk &     3532229 &        2954620 &     433798 &                297594 &    7 (5) &              467.79s &  $\specInt$\\
\bottomrule
\end{tabular}

\label{table:synth_stats}
\end{table*}

%% file: contents/6_related.tex
\section{Related Work}

\para{Pattern matching raising}
The raising of LLVM IR to a higher level MLIR has been investigated in MLT \cite{chelini2021progressive}. It develops a language that describes Affine IR \cite{moses2021polygeist} patterns and their corresponding replacement in Linalg IR. While flexible, it requires the writing of matching code for each pattern of interest. Furthermore, the replacement, or builder, code has to be rewritten for every target and is not scalable with IR evolution. A similar approach was investigated in \cite{ginsbach2018automatic} where an external constraint language is used to pattern match LLVM IR. Unlike MLT it replaces matches with calls to external APIs and again has to be rewritten for changing targets. 

\para{API replacement}
Replacing matched code/IR to a fixed API call \cite{collie2020m3} is a limited form of raising. KernelFarer \cite{de2021kernelfarer} works at the program level and restricts its attention to just GEMM API targets, but is more robust than IDL matching significantly more user code. This robustness is extended further in \cite{woodruff2022bind, martinez2023matching} which uses behavioral equivalence to match code. Such approaches, however, are intrinsically limited as they focus on fixed APIs rather than the open-ended nature of DSLs and their IRs. 

\para{Raising with synthesis}
Using program synthesis to generate programs from a specification is a long-studied area~\cite{fedyukovich2017gradual,singh2014modular}. Using a low-level program as the specification and a high level-one as the target was tacked in \cite{kamil2016verified}. Here appropriate stencil like loops in FORTRAN are lifted to their equivalent in Halide \cite{ragan2013halide}. This has been extended to a more generic LLVM framework \cite{ahmad2016leveraging} based on a common IR. While this has the potential to allow lifting to multiple targets \cite{ahmad2018automatically, ahmad2017optimizing}, it requires the compiler writer to provide a compiler and decompiler from each potential source and target into the IR which is not scalable. MLIR-Fuzz~\cite{mlirfuzz} offers a fuzzer \texttt{mlir-enumerate} which enumerates type-correct MLIR programs bottom-up for any dialect by translating MLIR's TableGen files into the dialect definition language IRDL~\cite{irdl}. 

\para{Example driven synthesis}
The use of input/output examples to synthesize high-level code has been explored in a number of projects~\cite{gulwani2011automating, zohar2018automatic,collie2020modeling,collie2021program}. It has been used to generate pytorch or tensor-flow code from tensor inputs~\cite{shi2022tf, nam2022predictive}. TF-coder~\cite{shi2022tf} uses type- constraints and equivalences to efficiently apply enumerative program search while \cite{nam2022predictive} uses a DeepCoder~\cite{balog2017deepcoder} style predictive model to guide code generation. AutoPandas~\cite{bavishi2019autopandas} uses a more powerful graph neural network based model to guide the generation of Panda code. As there is no ground-truth program to lift, just examples, such schemes cannot be directly used for IR raising. Furthermore, both source format and target output are hardwired for each domain. 

\para{Polyhedral compilation}
The use of polyhedral analysis to drive program optimization \cite{bondhugula2008pluto} has been extensively explored in the compilation community \cite{grosser2011polly}. It has been used for driving systolic code generation \cite{gross1986compilation}, memory hierarchy optimization, parallelization and GPU code generation \cite{baghdadi2019tiramisu} and forms the core for many modern tensor algebra compilers. Polly \cite{grosser2012polly} is able to generate efficient cache optimized and parallel code directly from LLVM IR. 

Low-level loop and memory reference representation in LLVM IR can make analysis difficult. This has motivated LLVM IR extensions to facilitate parallelization \cite{tian2017llvm} and motivated Polygeist, a C to Affine IR compiler \cite{moses2021polygeist} that uses Pluto \cite{bondhugula2008pluto} to generate cache optimized and parallel code. All of these approaches use polyhedral analysis to lower code, rather than mlirSynth which uses it to raise dialect levels. 

%% file: contents/7_conclusion.tex
\section{Conclusion and Outlook}
This paper presents a bottom-up enumerative synthesis approach to raising dialect levels within MLIR. The retargetable approach is applied to Affine IR, raising it to the Linalg and HLO IRs. It is applied to PolyBench and when the raised IR code is compiled to three platforms, it outperforms existing compilation flows. Future work will raise to programs with several target dialects, which needs a faster and more scalable synthesis algorithm. We plan to improve the synthesis search by re-using previous program space explorations, and the full integration of model checking into the synthesis process. We will also evaluate raising to new and emerging dialects of MLIR and apply to larger benchmark suites.